\documentclass[10pt]{wlscirep}
\usepackage{graphicx}% Include figure files
\graphicspath{{Figures/}}

\usepackage[utf8]{inputenc}
\usepackage[T1]{fontenc}
\title{Hybrid quantum-classical approach to enhanced quantum metrology}

\author[1,+]{Xiaodong Yang}
\author[1,2,+]{Xi Chen}
\author[3,4,*]{Jun Li}
\author[1,5,6,*]{Xinhua Peng}
\author[2,7,8]{Raymond Laflamme}
\affil[1]{Hefei National Laboratory for Physical Sciences at the Microscale and Department of Modern Physics, University of Science and Technology of China, Hefei 230026, China}
\affil[2]{Institute for Quantum Computing and Department of Physics and Astronomy, University of Waterloo, Waterloo N2L 3G1, Ontario, Canada}
\affil[3]{Shenzhen Institute for Quantum Science and Engineering and Department of Physics, Southern University of Science and Technology, Shenzhen 518055, China}
\affil[4]{Guangdong Provincial Key Laboratory of Quantum Science and Engineering, Southern University of Science and Technology, Shenzhen, 518055, China}
\affil[5]{CAS Key Laboratory of Microscale Magnetic Resonance, University of Science and Technology of China, Hefei 230026, China}
\affil[6]{Synergetic Innovation Center of Quantum Information and Quantum Physics, University of Science and Technology of China, Hefei 230026, China}
\affil[7]{Perimeter Institute for Theoretical Physics, 31 Caroline Street North, Waterloo, Ontario, N2L 2Y5, Canada}
\affil[8]{Canadian Institute for Advanced Research, Toronto, Ontario M5G 1Z8, Canada}

\affil[*]{lij3@sustech.edu.cn}
\affil[*]{xhpeng@ustc.edu.cn}

\affil[+]{these authors contributed equally to this work}

\begin{abstract}
Quantum metrology plays a fundamental role in many scientific areas. However, the complexity of engineering entangled probes and the external noise raise technological barriers for realizing the expected precision of the to-be-estimated parameter with given resources. Here, we address this problem by introducing adjustable controls into the encoding process and then utilizing a hybrid quantum-classical approach to automatically optimize the controls online. Our scheme does not require any complex or intractable off-line design, and it can inherently correct certain unitary errors during the learning procedure.  We also report the first experimental demonstration of this promising scheme for the task of finding optimal probes for frequency estimation on a nuclear magnetic resonance (NMR) processor. The proposed scheme paves the way to experimentally auto-search optimal protocol for improving the metrology precision.
\end{abstract}

\begin{document}

\flushbottom
\maketitle
% * <john.hammersley@gmail.com> 2015-02-09T12:07:31.197Z:
%
%  Click the title above to edit the author information and abstract
%
\thispagestyle{empty}

\section*{Introduction}
Measuring physical parameters of interest with highest precision remains the everlasting pursuit in science and technology \cite{GI11}. The general measurement procedure reads: prepare a probe, interact it with the system, and measure the probe. During this process, errors will result in a statistical uncertainty on the interested parameter $\phi$. These errors mainly come from intrinsic fluctuations, insufficient controls and external perturbations \cite{ED11,EDR11,YF17}. The central limit theorem tells us that repeated applications of this process $N$ times can improve the estimation precision, inducing a bound of $\Delta \phi \sim 1/\sqrt{N}$, which is called Standard Quantum Limit. Quantum metrology \cite{GV06,GI11,GT14,PS18} exploits available quantum resources to beat this limit and can approach a scaling called Heisenberg Limit, namely $\Delta \phi \sim 1/N$. However, in practical applications, realizing the expected precision under many cases, including inevitable external noise \cite{CH12,DR12,DR14}, complex probe states \cite{PCL12,JKF09,SJK10,MMJ04,RPP07,JJM11} and complicated encoding dynamics \cite{BF07,RB08,NK11,HW12}, are often very challenging.

 Fortunately, additional controls were found to be useful and necessary for quantum metrology to address these issues \cite{PJ17}. Dynamical decoupling methods \cite{TQH13,LJL15,PMW16} and quantum error corrections \cite{WMF14,KLS14} were used specifically to defend against certain external noise for maintaining the precision. For extended types of encoding dynamics, including time-dependent \cite{PJ17}, noncommuting \cite{HR19}, or general form \cite{YFC15}, carefully designed controls were applied to alter the dynamics and enhance the estimation precision. The above mentioned control methods are for specific purposes and are often very complex to design. Recently, Yuan and Liu \cite{LJY17} proposed a systematic controlled sequential scheme to search the required controls in noisy system for enhancing the quantum metrology abilities. It is based on an optimal control algorithm called Gradient Ascent Pulse Engineering (GRAPE), where the added controls could be iteratively refreshed until the performance function (e.g., quantum Fisher information \cite{GI11,PS18}) reaches the optimum. This algorithm is very efficient and easily-implemented for small-scale systems. However, in actual applications, it often happens that an exact model of the noise is lacking so that it is difficult to evaluate the gradient of the performance function to a good precision, even for the single-qubit case. These problems are further harmed by the exponentially increased complexity of the system dynamics. 
 
 To tackle these issues, we utilize a hybrid quantum-classical approach \cite{LY17,LL17} assisted GRAPE (hqc-GRAPE for short) to practically learn the optimal controls experimentally. Under a completely different motivation, the previous works concern how to speedup quantum optimal control problems, while here we seek for its extension to quantum metrology area. Hybrid quantum-classical (HQC) algorithms, which combine the present-day accessible quantum resources with sophisticated classical computation routines, have witnessed tremendous successful applications, ranging from simulating quantum chemistry \cite{PM14,GE19,WH15,OP16} to solving optimization problems \cite{LY17,LL17,BW16,SC19}. By applying this approach, we do not require any prior-knowledge of how the optimal controls are related to the encoding dynamics, as they are automatically learned in the experiments without any design. These searched controls and the encoding dynamics are then coupled together to deliver an optimal metrology procedure. The computationally resource-consuming and experimentally intractable parts of the GRAPE algorithm, namely the gradient of the performance function, are efficiently and conveniently measured by applying some single-qubit rotations to the system. Furthermore, as this HQC approach is combined with GRAPE to deliver a closed-loop learning \cite{BC10} procedure, it has inherent features of defending against certain kinds of unitary noise for improving the metrology precision.
 
 We also presented a demonstrative experiment of finding optimal probes for estimating the frequency by hqc-GRAPE on a two-qubit NMR processor. The experimental results verify the success of hqc-GRAPE in learning optimal controls for improving the metrology precision. The outline of this study is described as follows. Firstly, we introduce the details of hqc-GRAPE for quantum metrology in ``Framework'' section. The experimental procedure is presented in ``Experiment'' section. Finally, we provide some conclusions and discussions in ``Discussion'' section.

\section*{Results}
\noindent 
\textbf{\large{Framework}}\\
Consider a typical quantum metrology task of estimating an interested parameter $\Omega$ which is encoded in a general form of Hamiltonian $\mathcal{H}_0(\Omega)$ (with couplings between system qubits). Conventional quantum metrology schemes then proceed to design optimal probe state and the corresponding optimal measurements to gain the best metrology precision. In particular, the metrology can be thought of as two distinct tasks: (1) Find a classical procedure that enables us to engineer the probe state whose quantum Fisher information is sufficiently optimal. (2) Application of the encoding process to an optimal probe (synthesized using the above procedure), then estimate the interested parameter with suitable measurements. Here, we mainly focus on the first task. In practice, inevitable noise and the complexity of synthesizing the probes will prevent us from realizing the best precision. As stated in the introduction part, additional controls can be applied to address these problems and improve the precision that can be reached.
 
\begin{figure}
\centering
\includegraphics[width=0.48\textwidth,height=0.28\textwidth]{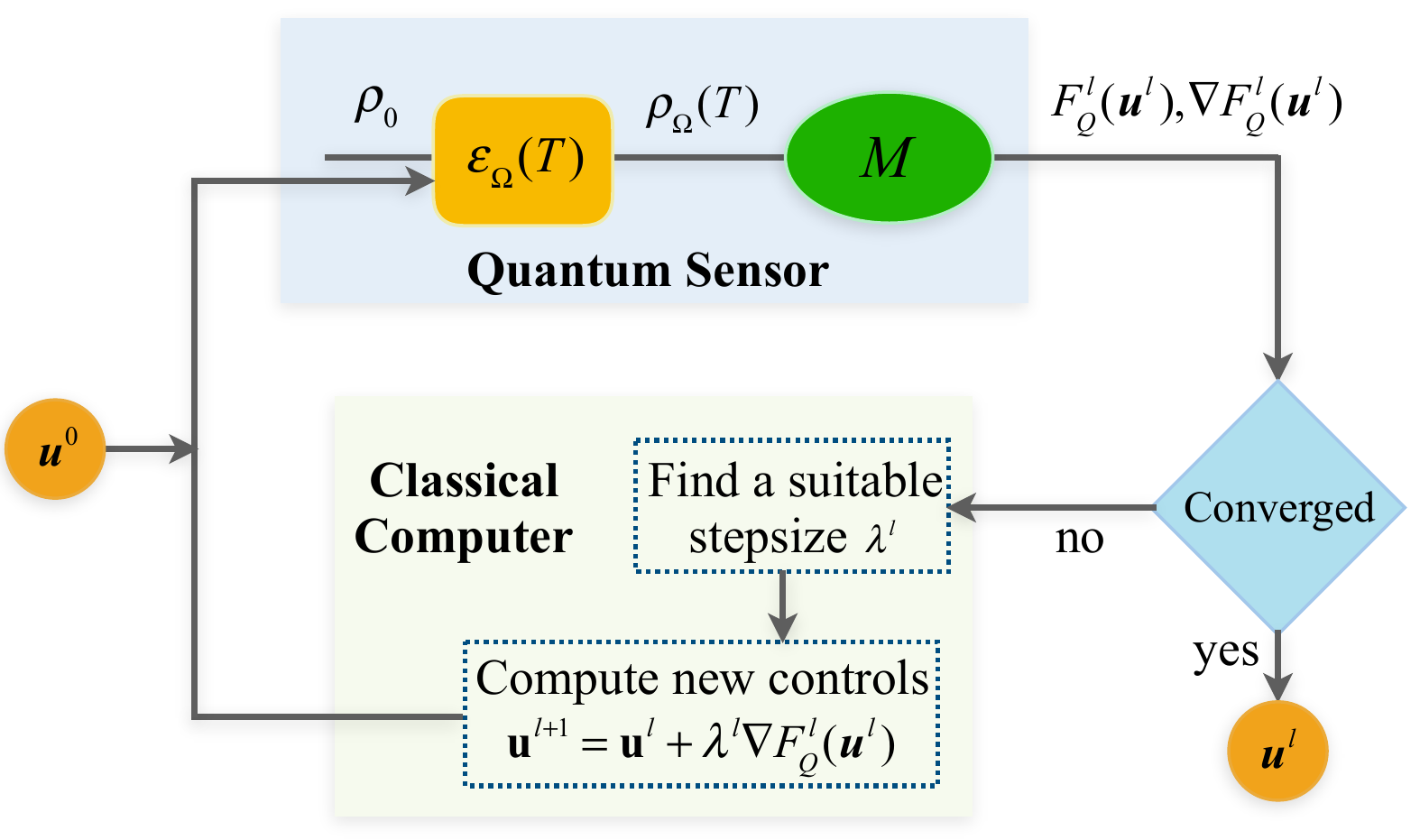}
\caption{Schematic diagram of hybrid quantum-classical approach assisted GRAPE (hqc-GRAPE) for quantum metrology. From easily prepared state $\rho_0$, initial controls $\mathbf{u}^0$ is imported into the encoding process, resulting in $\varepsilon_\Omega(T)$ over an encoding time $T$. The resultant state $\rho_\Omega(T)$ is then measured to obtain the quantum Fisher information $F_Q^l(\mathbf{u}^l)$ and its gradient $ \nabla F_Q^l(\mathbf{u}^l)$ in $l$-th iteration. Afterwards, a suitable stepsize $\lambda_l$ is determined to generate new controls by ${{\bf{u}}^{l + 1}} = {{\bf{u}}^l} + {\lambda ^l}\nabla F_Q^l({\mathbf{u}^l})$, which are imported into the encoding process for the next iteration $l+1$. This procedure is looped until the stopping criterion is met. Here, the quantum sensor is combined with the classical computer to deliver a practical hybrid approach for quantum metrology.
}
\label{scheme}
\end{figure}

Here, we implement adjustable controls to alter the encoding dynamics, thus the total Hamiltonian can be expressed as 
\begin{equation}
	\mathcal{H}
	=\mathcal{H}_0(\Omega)+\sum_{k=1}^K u_k(t)\mathcal{H}_k,
\end{equation}
where $u_k(t)~(t\in(0,T))$ represents the amplitude of the $k$-th control field. Note that $\mathcal{H}_0(\Omega)$ contains the interactions between qubits, thus the above total Hamiltonian captures a general form for the metrology application. Without loss of generality, we assume the $k$-th control Hamiltonian can be formulated as $\mathcal{H}_k=\sigma_\alpha^k$ with $\sigma_\alpha^k~(\alpha=x,y,z)$ being the Pauli matrix, i.e., the controls are at most three directions for each qubit, which is a standard form in many quantum systems \cite{VLC05,MYS01,LDB03}. To optimize these controls by hqc-GRAPE, we divide the total evolution time $T$ into $M$ equal segments, and the controls in each segment of duration $\Delta t=T/M$ are approximately treated as constants. Thus, in the $m$-th segment, the sliced evolution operator can be depicted by 
\begin{equation}
	\varepsilon_\Omega^m=\text{exp}\left\{-i \Delta t \left[ \mathcal{H}_0(\Omega)+\sum_{k=1}^K u_k[m]\mathcal{H}_k \right] \right\},
\end{equation} 
this will lead to the total evolution operator $\varepsilon_\Omega(T)=\prod_{m=1}^M \varepsilon_\Omega^m$. 

The metrology process using hqc-GRAPE begins with some easily prepared pure probe state $\rho_0$, which does not need to be optimal. This avoids the complex design and synthesis of the optimal probes, thus greatly easing the analytical efforts. The probe $\rho_0$ is then engineered by the system evolution with some trial control fields $\mathbf{{u}}=(u_k[m]),k=1,2,...,K;m=1,2,...,M$, leading to the final system state $\rho_\Omega(T)=\varepsilon_\Omega(T) \rho_0 \varepsilon_\Omega(T)^\dag$. Performing the corresponding optimal measurements will induce the best metrology precision that can be reached in this situation. Here, to quantify the performance of estimating the interested parameter $x$, we can use the quantum Fisher information ($F_Q$ for short) as a performance function \cite{TP13,AO17}, namely  
\begin{equation}
	F_Q(\Omega,\mathbf{u})=4T^2\left\{ \text{Tr}[\rho_\Omega(T) (\partial_\Omega \mathcal{H}_0)^2]-\text{Tr}[\rho_\Omega(T) \partial_\Omega \mathcal{H}_0]^2 \right\}.
	\label{Eq.Fq}
\end{equation}
In order to achieve the possibly best precision with the given resources, we need to iteratively refresh the control fields to maximize the performance function $F_Q$. In hqc-GRAPE, the control fields are updated by moving towards the gradient direction of the performance function with some appropriate distance.  The explicit form of the gradient of $F_Q$, i.e., $\nabla F_Q=\mathbf{g}=(g_k[m])$ with $g_k[m]={\partial F_Q}/{\partial {u_k[m]}}$, can be easily calculated as follows
\begin{eqnarray}\label{gradient}
	g_k[m] 
	&=& 4T^2\left\{ \text{Tr} \left[ \frac{\partial \rho_\Omega(T)}{\partial {u_k[m]}} (\partial_\Omega \mathcal{H}_0)^2 \right] \right. \\ \nonumber
	&& \left. - 2 \text{Tr} \left[ \frac{\partial \rho_\Omega(T)}{\partial {u_k[m]}} \partial_\Omega \mathcal{H}_0 \right] \text{Tr} \left[  \rho_\Omega(T) \partial_\Omega \mathcal{H}_0 \right] \right\},
	\label{Eq g}
	\end{eqnarray} 
For brevity, we denote $U_{m_1}^{m_2}=\varepsilon_\Omega^{m_2} \cdots \varepsilon_\Omega^{m_1+1} \varepsilon_\Omega^{m_1}$, then we get ${\partial \rho_\Omega(T)}/{\partial {u_k[m]}}=-i \Delta t U_{m+1}^M [\sigma_\alpha^k,U_1^m \rho_0 {U_1^{m}}\dag] {U_{m+1}^{M}}^\dag$. The key idea of hqc-GRAPE is that we can compute this commutator by some local rotations \cite{LY17}. It is achieved by using the relation which holds for any operator $\rho$ 
\begin{equation}
	\left[\sigma_{\alpha}^{k}, \rho\right]=i\left[R_{\alpha}^{k}\left(\frac{\pi}{2}\right) \rho R_{\alpha}^{k}\left(\frac{\pi}{2}\right)^{\dagger}-R_{\alpha}^{k}\left(-\frac{\pi}{2}\right) \rho R_{\alpha}^{k}\left(-\frac{\pi}{2}\right)^{\dagger}\right],
\end{equation} 
where $R_\alpha^k(\pm \pi/2)$ represents the $\pm \pi/2$ rotations along $\alpha$ axis. Thus, we can get 
\begin{eqnarray}\label{rhott}
	\frac{\partial \rho_\Omega(T)}{\partial {u_k[m]}}
	=\Delta t \left\{ U_{m+1}^{M} R_{\alpha}^{k}(\frac{\pi}{2}) U_{1}^{m} \rho_{0}\left[ U_{m+1}^{M} R_{\alpha}^{k}(\frac{\pi}{2}) U_{1}^{m} \right]^\dag \right.\\ \nonumber
	\left. - U_{m+1}^{M} R_{\alpha}^{k}(-\frac{\pi}{2}) U_{1}^{m} \rho_{0}\left[U_{m+1}^{M} R_{\alpha}^{k}(-\frac{\pi}{2}) U_{1}^{m}\right]^\dag  \right \}.
\end{eqnarray}
In this way, by inserting local rotations, we can obtain the $m$-th gradient information similarly as presented in Eq. \ref{Eq.Fq}, i.e., directly measuring the $F_Q$ of the final system state involving the inserted local rotations. Note that this transformation does not depend on how we measure the $F_Q$. Thus, $2KM$ operations are needed to compute the gradient $\mathbf{g}$ in each iteration.

Overall, one needs $2KM+1$ measurements of the performance function in each iteration. In general, $K$ scales polynomially with the increasing of qubits, as the control Hamiltonians $\mathcal{H}_k$ are single Pauli matrixes along at most three directions for each qubit. Typically, $M$ increases polynomially with the growing of system size. Indeed, for most randomly selected Hamiltonian $\mathcal{H}$, the minimal number of the controls required to synthesize it will scale exponentially. However, near-term quantum metrology applications are likely concerned with what can be done with a polynomial number of gate operations. This corresponds to optimizing over the best possible probes that can be synthesized with a polynomial number of control slices -- which is precisely the problem our protocol is ideally suited for. Thus, for the practicality of our protocol, the key issue becomes how to efficiently measure the performance function, i.e., $F_Q$. Fortunately, there have emerged several scalable methods to estimate $F_Q$ in experiment, where they have replaced $F_Q$ with some easily accessible quantities, such as (1) purity loss \cite{modi2016fragile,yang2020probe}. $F_Q$ is bounded by purity loss, which captures how fragile the purity of the resultant state with respect to stochastic noise on the encoding parameter and can be obtained by simulating a finite stochastic noise regardless of the system size. 
(2) multiple-quantum coherence (MQC)\cite{GMH18}. By appending reversion of the system dynamics, MQC can be efficiently accessed and used to calculate $F_Q$ in an experiment. This procedure takes finite runs of experiments for Fourier transformation of the measured signal, thus does not   need exponential resource.
(3) Loschmidt echo \cite{MTS16}. This method is similar as method (2) but needs added controlled operations and an ancillary qubit. However, it carries a great advantage of readout from a single ancillary qubit. In real experiment, it is advisable to choose the suitable method in consideration of the experimental resource needed.

We proceed by briefly summarizing the algorithmic procedure of hqc-GRAPE (see schematic diagram in Fig. \ref{scheme}) for solving this metrology task: \\
\textit{Step 1}: Randomly generate initial control field $\mathbf{u}^0$, then apply it to some easily prepared probe state $\rho_0$. The system state will evolve under this control together with the encoding dynamics governed by $\mathcal{H}_0$. Measure the performance function $F_Q^0(\mathbf{u}^0)$ and the corresponding gradient $\mathbf{g}^0=\nabla F_Q^0(\mathbf{u}^0)=(g_k^0[m]),k=1,2,...,K; m=1,2,...,M$.\\
\textit{Step 2}: Set the iteration number as $l=l+1$, calculate the updated controls by $\mathbf{u}^{l+1}=\mathbf{u}^{l} +\lambda^l \mathbf{g}^l$,where $\lambda^l$ is some appropriate stepsize along the gradient direction and $\mathbf{g}^l=\nabla F_Q^l(\mathbf{u}^l)=(g_k^l[m])$. Measure the performance function $F_Q^{l+1}$ and the gradient $\mathbf{g}^{l+1}$ again.\\
\textit{Step 3}: Check whether the measured performance function satisfies the stopping criterion, if not, go to \textit{Step 2}.

In this closed-loop learning procedure, the resource-consuming parts, i.e., the computing of $F_Q$ and its gradient $\nabla F_Q$, are efficiently accomplished by the quantum system. The classical computer is used to determine the suitable stepsize for updating controls fields, to generate the pulses for single-qubit rotations, and to store the data in each iteration. The resources needed for the classical computer are then very moderate, even for very large quantum systems. Therefore, the cooperated scheme of quantum sensor and classical computer is very applicable for the near-term quantum metrology tasks with accessible resources.

\bigskip
\noindent
\textbf{\large{Experiment}}\\
\textbf{Setup and techniques.}
The proof-of-principle experiments were conducted using the $^{13}\text{C}$-labeled sample Chloroform on a Bruker Avance III 400 MHz spectrometer at room temperature. We mark the spins $^{13}\text{C}$, $^{1}\text{H}$ as 1 and 2, respectively. The internal Hamiltonian can be described as ${\mathcal{H}_{{\mathop{\rm int}} }} =  \sum\nolimits_{i = 1}^2 {\Omega^i \sigma_z^i}/2  + \pi J\sigma_z^1 \sigma_z^2/2$, where $\Omega^i$ represents the offset of the $i$-th spin in the rotating frame and $J=214.5~$Hz is the scalar coupling strength between the two spins.

For brevity, we set $\Omega^1=\Omega^2=\Omega$ and consider estimating the single parameter $\Omega$ encoded in the following Hamiltonian $\mathcal{H}_0 (\Omega)=\Omega(\sigma_z^1+\sigma_z^2)/2  + \pi J\sigma_z^1 \sigma_z^2/2$. Additional control fields  are introduced along $x$ and $y$ directions of each spin, thus leading to $\mathcal{H}=\mathcal{H}_0(\Omega)+\sum ^{2}_{k=1}\left( u_{k,x}[m]\sigma_x^k+u_{k,y}[m]\sigma_y^k\right)$. For an encoding time $T$, the to-be-optimized control fields are sliced into $M$ segments with $\mathbf{u}=(u_{k,x}[m],u_{k,y}[m])$, where $k=1,2;m=1,2,...,M$. In this simple case, we do not need to seek for advanced methods as mentioned above to estimate $F_Q$. Specifically, the quantum Fisher information of the resultant state $\rho_\Omega(T)$ corresponding to the controls $\mathbf{u}$ can be explicitly written as $F_Q(\mathbf{u})=T^2 \{ \text{Tr}[\rho_\Omega(T)(\sigma_z^1+\sigma_z^2)^2] - \text{Tr}[\rho_\Omega(T)(\sigma_z^1+\sigma_z^2)]^2 \} $. Note that the trace operations only concern the diagonal elements, and the Pauli matrix $\sigma_z^k$ is diagonal, thus only the diagonal elements of $\rho_\Omega(T)$ matter. Direct derivation indicates that only two diagonal elements of $\rho_\Omega(T)$ are needed to compute $F_Q(\mathbf{u})$, which greatly reduces the experimental cost. Similarly, the gradient of $F_Q(\mathbf{u})$ reduces to $g_{k,\alpha}[m]=T^2 \{ \text{Tr}[\rho'(T)(\sigma_z^1+\sigma_z^2)^2] - 2\text{Tr}[\rho'(T)(\sigma_z^1+\sigma_z^2)]  \text{Tr}[\rho_\Omega(T)(\sigma_z^1+\sigma_z^2)]  \} $ with $\rho'(T)={\partial \rho_\Omega(T)}/{\partial {u_{k,\alpha}[m]}}$ and $\alpha=x,y$, where $\rho'(T)$ is obtained by applying local rotations on the $k$-th spin during the $m$-th sliced controls.

\begin{figure*}
\centering
\includegraphics[width=0.8\textwidth,height=0.54\textwidth]{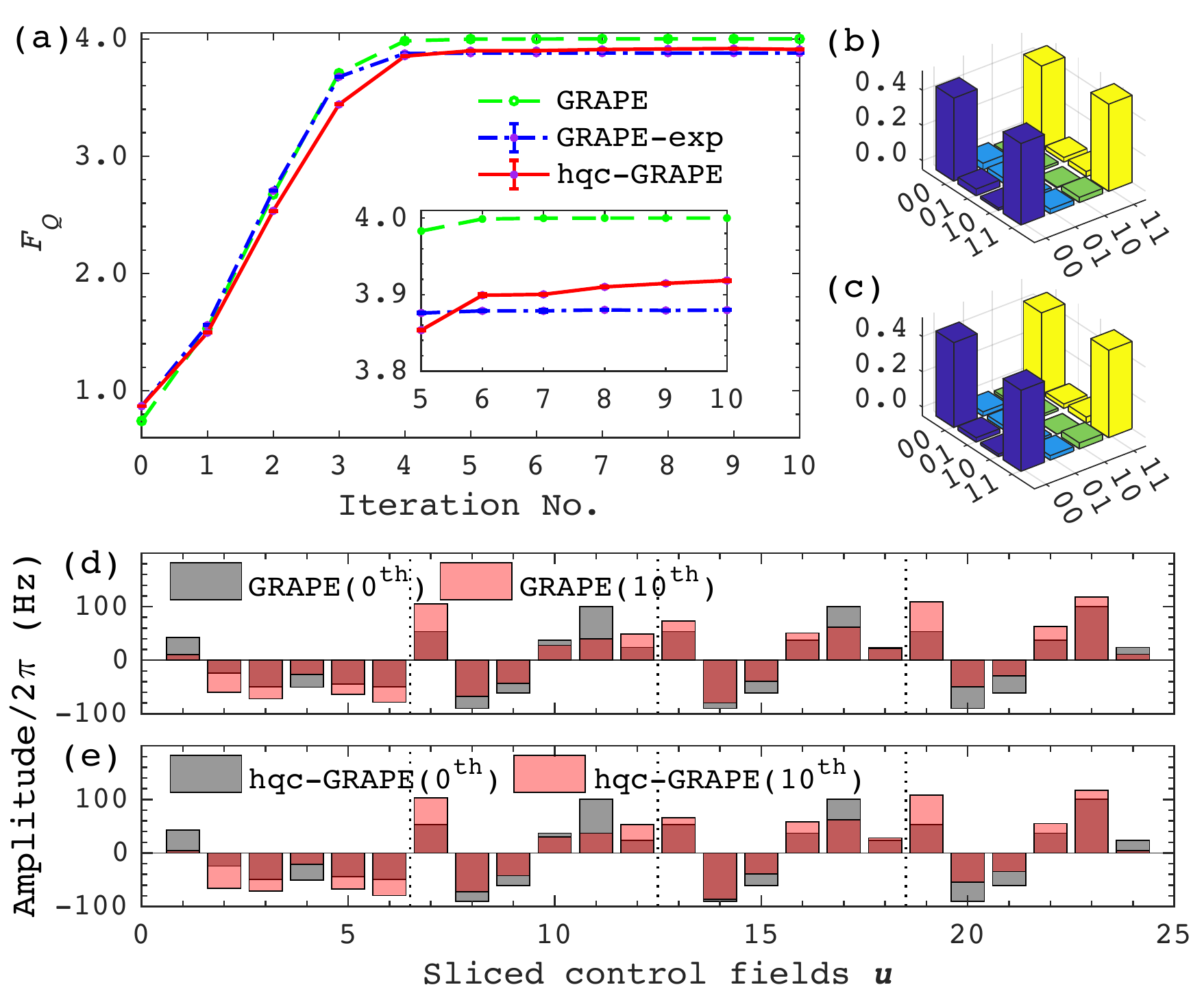}
\caption{Experimental results of GRAPE-exp and hqc-GRAPE, together with simulation results of GRAPE for quantum metrology. 
(a) shows the optimization process of GRAPE (green dashed line), GRAPE-exp (blue dash-dotted line) and hqc-GRAPE (red solid line). In each iteration, $F_Q$ is measured five times to induce the statistical error bars. The subplot enlarges the detailed performance of these three approaches close to convergence. 
(b) and (c) illustrate the amplitude of the density matrix of one typical optimal resultant state $\rho_\Omega(T)$ obtained by GRAPE-exp and hqc-GRAPE, respectively. 
(d) and (e) plot the amplitude of the initial control fields (grey bars) and the final optimal control fields (pink bars) for GRAPE and hqc-GRAPE, respectively. The amplitude overlapping parts of the initial controls and the final controls are brown.}
\label{exp}
\end{figure*}

\noindent
\textbf{Experimental procedures and results.}
The detailed experimental procedure of hqc-GRAPE can be divided into the following five steps: 

(i) \textit{Preparation of initial state $\rho_0$.} We initialized the system at pseudo-pure (PPS) state by line-selective method \cite{peng2001preparation}, i.e., ${\rho_{\text{pps}}} = \frac{{1 - \varepsilon }}{{4}}{\mathbf{I}_{4}} + \varepsilon |00\rangle \langle00|$, where $\mathbf{I}_{4}$ represents the $4 \times 4$ identity matrix and $\varepsilon  \approx {10^{ - 5}}$ is the thermal polarization of the two-qubit system. Notice that the identity matrix doesn't produce observable effects, thus the initial PPS state  effectively behaves like $\rho_0^{th}=|00\rangle \langle00|$. Full tomography \cite{lee2002quantum} verified that the prepared initial state $\rho_0$ has a fidelity of 0.9986 compared with $\rho_0^{th}$ by defining $F(\rho_0^{th},\rho_0)=\text{Tr}(\rho_0^{th}\rho_0)/\sqrt{\text{Tr}[(\rho_0^{th})^2]\text{Tr}[(\rho_0)^2]}$.
 
(ii) \textit{Generation of initial controls $\mathbf{u^{0}}$.} The initial control fields $\mathbf{u}^0=(u_{k,x}^0[m],u_{k,y}^0[m])$ with $k=1,2; m=1,2,...,M$ were randomly generated on classical computer and applied to the quantum simulator. During the optimization procedure, we set $\Omega=2\pi \times 50~$Hz, $M=6$ and the encoding time $T=9~$ms. 

(iii) \textit{Measurement of $F_Q^l(\mathbf{u}^l)$ and $\mathbf{g}^l$.} In the $l$-th iteration, we first measured the performance of the resultant state $\rho_\Omega(T)$ corresponding to $\mathbf{u}^l$, namely $F_Q^l(\mathbf{u}^l)$. As stated above, only two diagonal elements of $\rho_\Omega(T)$ are needed to compute its quantum Fisher information. In our NMR simulator, this was accomplished by applying two local $\pi/2$ rotations along $y$ axis on spin 1 and 2, respectively, and observing the corresponding spectra \cite{lee2002quantum}. To obtain $\mathbf{g}^l=(g_{k,x}^l[m],g_{k,y}^l[m])$, after the $m$-th sliced evolution operator, we inserted two groups of local rotations $R_x^k(\pm\pi/2)$ and $R_y^k(\pm\pi/2)$ sequentially and measured the resultant state $\rho'(T)={\partial \rho_\Omega(T)}/{\partial {u_{k,\alpha}[m]}}, \alpha=x,y$ according to Eq. \ref{gradient} and Eq. \ref{rhott}. Similarly, only diagonal elements of $\rho'(T)$ are necessary to compute $\mathbf{g}^l$.
  
(iv) \textit{Generation of new controls $\mathbf{u}^{l+1}$.} The measured $F_Q^l(\mathbf{u}^l)$ and $\mathbf{g}^l$ were then fed back to the classical computer. A suitable stepsize $\lambda^l$ was decided to generate new controls by $\mathbf{u}^{l+1}=\mathbf{u}^{l} +\lambda^l \mathbf{g}^l$. Here, $\lambda^l$ was initially set as 5000 and gradually decreased by $50\%$ if $F_Q^l(\mathbf{u}^{l+1})$ was worse than $F_Q^l(\mathbf{u}^{l})$. 

(v) \textit{Loop of the optimization procedure.} The iteration number was set as $l=l+1$ and the refreshed controls $\mathbf{u}^{l+1}$ were applied to the NMR simulator again. We then jumped to step (iii) to loop the rest steps. This iterative procedure was stopped until the settled maximum iteration number 10 was hit.

Furthermore, in order to demonstrate the advantages of hqc-GRAPE in searching optimal protocol for quantum metrology in realistic experiments, we compared it with the conventional open-loop designs entirely running on classical computer, which we marked as GRAPE. This pure classical simulation iteratively calculates $F_Q^l(\mathbf{u}^l)$ and $\mathbf{g}^l$ according to the ideal Hamiltonian, which does not include the effects of the inevitable noises in real situation, thus deserves the above mentioned closed-loop optimization. We also directly applied the classically searched controls by GRAPE to the NMR quantum simulator to measure the corresponding $F_Q$ in each iteration, which was denoted as GRAPE-exp. Specifically, we first prepare the system at its initial state $\rho_0$ as described in the step (i). Next for each iteration $l$, we directly import the corresponding optimal controls searched by the open-loop GRAPE into the quantum simulator. Finally we measure its performance function as introduced in the step (iii).

The experimental results are shown in Fig. \ref{exp}. Green dashed line in Fig. \ref{exp}(a) shows the optimization process entirely running on classical computer by the GRAPE method. The searched optimal controls induce a final $F_Q$ of 4.00, which saturates the theoretical bound \cite{PS18}. Blue dash-dotted line demonstrates the measured $F_Q$ through directly applying the controls searched by the open-loop GRAPE to the NMR simulator. The final optimal controls are tested 5 times to estimate the statistical error resulting in $F_Q=3.8798 \pm 0.0006$ for GRAPE-exp. The deviation from the theoretical optimum attributes to various noise and errors existing in the control process. Red solid line then presents the optimization process of hqc-GRAPE, the final $F_Q$ corresponding to the learned optimal controls is $3.9102 \pm 0.0007$ over 5 tests. This indicates that in the metrology process, hqc-GRAPE method automatically corrected some forms of errors, thus reaching a higher $F_Q$ than that of the open-loop designs. We also reconstructed the final optimal resultant state $\rho_\Omega(T)$, as shown in Fig. \ref{exp}(b) for GRAPE-exp, and Fig. \ref{exp}(c) for hqc-GRAPE. Compared them with the theoretical optimal NOON state $\rho_t=|\psi_t\rangle\langle \psi_t|$,$|\psi_t \rangle =(|00\rangle +e^{i\phi}|11\rangle)/\sqrt{2}$ \cite{PS18}, we obtained a fidelity of $0.9954 \pm 0.0002$ for GRAPE-exp and $0.9962 \pm 0.0001$ for hqc-GRAPE by defining $F(\rho_t,\rho_\Omega(T))=\text{Tr}(\rho_t\rho_\Omega(T))/\sqrt{\text{Tr}[(\rho_t)^2] \text{Tr}[(\rho_\Omega(T)^2]}$. This reveals that hqc-GRAPE can reach a state closer to the theoretical optimum than GRAPE-exp does. Moreover, we plot the initial controls (0-th iteration) and the final optimal controls (10-th iteration) searched by GRAPE and hqc-GRAPE in Fig. \ref{exp}(d) and Fig. \ref{exp}(e), respectively. These two approaches started from the same initial controls, but terminated with slightly different control amplitudes, which leads to their distinct performances.

\begin{figure*}
\centering
\includegraphics[width=0.7\textwidth,height=0.48\textwidth]{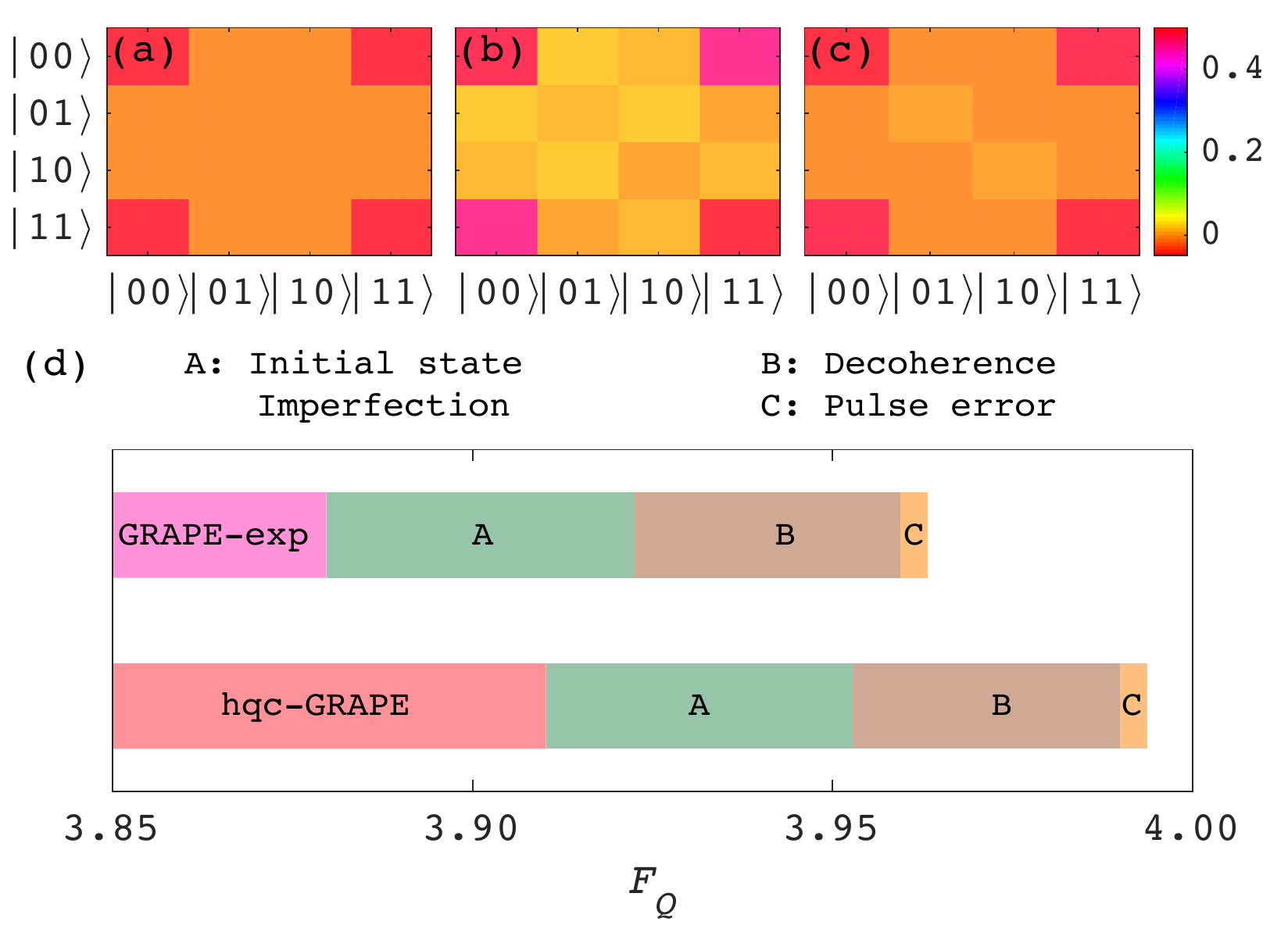}
\caption{Error analysis for the optimal resultant state $\rho_\Omega(T)$ and its $F_Q$. (a) shows the amplitude of the density matrix $\rho_\Omega(T)$ obtained by GRAPE. (b) and (c) then illustrate the results when considering initial state imperfection and decoherence further, respectively. (d) plots the final optimal $F_Q$ of GRAPE-exp and hqc-GRAPE. Moreover, three main forms of errors in the control process are gradually accumulated to demonstrate the error compensation results of GRAPE-exp and hqc-GRAPE.  }
\label{error}
\end{figure*}

\bigskip
\noindent
\textbf{\large{Analysis of the benefits of hqc-GRAPE}}

As demonstrated above, the final optimal controls searched by hqc-GRAPE induce a higher quantum Fisher information $F_Q$ than GRAEP-exp did. The benefits come from the inherent features of closed-loop learning, which can automatically correct some specific unitary errors \cite{BC10}. To explicitly understand how hqc-GRAPE improve the estimation precision, we now proceed to carefully analyze the existing errors in our experiments. In general, they can be divided into the following four types:

\textit{Initial state imperfection.} It refers to the deviation of experimentally prepared initial state and the desired one. From the experimentally reconstructed initial state $\rho_0$, we applied optimal controls searched by GRAPE, leading to the resultant state shown in Fig. \ref{error}(b) (the ideal resultant state is depicted in Fig. \ref{error}(a)). We then performed ideal measurements and got $F_Q=3.9573$. This indicates that initial state imperfection yields an error of 0.0427 from the theoretical optimal value. Actually, as the prepared $\rho_0$ is not a perfect pure state, the non-unitary parts under the controls and the encoding process will finally induce the errors on estimating $F_Q$. In addition, the spectrum of the prepared initial state is directly treated as the reference signal for characterizing $F_Q$. Thus, this kind of error will eventually cause non-unitary effects that can not be corrected in the closed-loop learning process.

\textit{Decoherence.} Normally, the effects of decoherence in NMR simulator can be described by phase damping channel $\varepsilon_{\text{PD}}$ and generalized amplitude damping channel $\varepsilon_{\text{GAD}}$ \cite{vandersypen2001experimental}. In each sliced evolution process, phase damping error was involved by $\rho \rightarrow \varepsilon_{\text{PD}} ^2  \circ \varepsilon_{\text{PD}}^1(\rho) $, where $\varepsilon_{\text{PD}} ^i(\rho)=(1-p_i)\rho+p_i\sigma_z^i \rho \sigma_z^i$ with $p_i=(1-e^{-\Delta t/T_2^i})/2, i=1,2$ being the qubit number, and $T_2^1=0.3~\text{s}, T_2^2=3.3~\text{s}$. Similarly, generalized amplitude damping error was expressed as $\rho \rightarrow \varepsilon_{\text{GAD}} ^2  \circ \varepsilon_{\text{GAD}}^1(\rho) $ and calculated by $\varepsilon_{\text{GAD}}^i(\rho)=\sum_s E_s^i\rho E_s^{i\dag}$, where 
\begin{eqnarray*}
E_0^i &=&\sqrt{p}\left(\begin{array}{cc} 1 & 0 \\ 0 & \sqrt{1-\eta^i} \end{array}\right), 
E_1^i =\sqrt{1-p}\left(\begin{array}{cc} 0 & 0 \\ \sqrt{\eta^i} & 0 \end{array}\right), \\
E_2^i &=&\sqrt{1-p}\left(\begin{array}{cc} \sqrt{1-\eta^i} & 0 \\ 0 & 1 \end{array}\right), 
E_3^i = \sqrt{p}\left(\begin{array}{cc} 0 & \sqrt{\eta^i} \\ 0 & 0 \end{array}\right), 
\end{eqnarray*}
with $\eta^i=1-e^{-\Delta t/T_1^i}, p \approx 1/2$ and $T_1^1=18.5~\text{s},T_1^2=9.9~\text{s}$. With perfect initial state, optimal pulses searched by GRAPE and ideal measurements, the decoherence then induces an error of 0.0370, as shown in Fig. \ref{error}(c). It's worth noting that the coherent controls may partially ease the decoherence \cite{PMW16}. However, the analysis above has taken the effects of the coherent controls into consideration, thus the remaining error can not be corrected.  

\textit{Pulse error.} To estimate the influence of pulse errors on $F_Q$, we assume that the amplitude of the controls undergoes uniformly distributed stochastic fluctuation with at most 5\% distortions. With perfect initial state and ideal measurements, we repeated the optimal controls with fluctuations 1000 times and got an error around 0.0038. 

\textit{Measurement error.} Measurement errors can be estimated from the stochastic fluctuations of NMR spectra. In our experiments, measurements are accomplished by observing the NMR spectra and fitting them with Lorentzian functions. A direct estimation of the measurement error of resultant state was at the level of $10^{-4}$, which becomes $10^{-6}$ when considering its $F_Q$ using error propagation. Reasonably, this type of error can be ignored.

To conclude, initial state imperfection, decoherence and pulse error are three major errors in our experiments. However, as analyzed above, the initial state imperfection here will cause non-unitary effects and the error of decoherence here is the part that controls can not handle. That is to say, the employed coherent controls are not able to further deal with these two errors. Thus when we compensate the loss of these two errors on $F_Q$, the performance of hqc-GRAPE is remarkably improved to 3.9899, which is near the optimal value, as shown in Fig. \ref{error}(d). For the results of GRAPE-exp after error compensation, there is still a visible gap with respect to the optimal value, see Fig. \ref{error}(d). These results indicate that hqc-GRAPE can intelligently correct pulse error and some other unknown unitary errors to improve the metrology precision.

\section*{Conclusion}

For quantum metrology, additional controls are helpful for dealing with the challenges of external noise, complicated designs and manipulation of probes and encoding dynamics. Though this area has attracted much attention recently, practical schemes are still urgently in demand \cite{yang2020probe}. In this study, we proposed a hybrid quantum-classical approach assisted GRAPE to automatically engineer the encoding dynamics for searching optimal probes to improve the metrology precision. The quantum simulator, which can efficiently simulate the time-consuming part of the GRAPE algorithm, is combined with the classical computer to iteratively optimize the controls. In our scheme, there is no need to start from optimal probes, the controls can transform arbitrary pure initial probe to the best resultant state during the learning process without any prior designs. Furthermore, many specific unitary errors can be inherently corrected by this closed-loop learning procedure, which indeed improve the metrology precision. The accompanied experiments successfully verified the effectiveness and advantages of hqc-GRAPE. 

The demonstrative experiments were implemented on a small-scale NMR quantum simulator, however, the proposed scheme is scalable and feasible for current NISQ \cite{preskill2018quantum} systems. Cooperated with many efficient methods of estimating quantum Fisher information in experiments \cite{modi2016fragile,GMH18,MTS16}, the proposed scheme is promising in realizing optimal quantum metrology with auto-design techniques for more complicated and large-sized applications.

%\bibliography{Reference.bib}
\providecommand{\noopsort}[1]{}\providecommand{\singleletter}[1]{#1}%

\section*{Acknowledgements}

J.L. is supported by the National Natural Science Foundation of China (Grant No. 11975117), and Guangdong Provincial Key Laboratory (Grant No. 2019B121203002). X.P. is supported by National Key Research and Development Program of China (Grant No. 2018YFA0306600), the National Science Fund for Distinguished Young Scholars (Grant No. 11425523), Projects of International Cooperation and Exchanges NSFC (Grant No. 11661161018), and Anhui Initiative in Quantum Information Technologies (Grant No. AHY050000). 

\section*{Author contributions statement}

X.Y. proposed this project and wrote the manuscript; X.C. and X.Y. performed the experiments and analyzed the data; J.L. and R.L. helped with discussions and revised the manuscript; X.P. supervised the whole project.

\section*{Conflict of interest}
The authors declare that they have no conflict of interest.

\section*{Data availability}
The datasets generated during and/or analysed during the current study are available from the corresponding author on reasonable request.

\end{document}